\documentclass[sigconf, natbib=true]{acmart}
\usepackage{multirow}
\AtBeginDocument{%
  \providecommand\BibTeX{{%
    \normalfont B\kern-0.5em{\scshape i\kern-0.25em b}\kern-0.8em\TeX}}}

\setcopyright{acmcopyright}
\copyrightyear{2018}
\acmYear{2018}
\acmDOI{XXXXXXX.XXXXXXX}

\acmConference[Conference acronym 'XX]{Make sure to enter the correct
  conference title from your rights confirmation emai}{June 03--05,
  2018}{Woodstock, NY}
\acmPrice{15.00}
\acmISBN{978-1-4503-XXXX-X/18/06}

\begin{document}

\title{PALR: Personalization Aware LLMs for Recommendation}

\author{Fan Yang}
\authornote{Authors contributed equally to this research. Names are ordered alphabetically.}
\email{ffanyang@amazon.com}
\orcid{1234-5678-9012}
\affiliation{%
  \institution{Amazon Alexa AI}
  \city{Seattle}
  \state{WA}
  \country{USA}
}

\author{Zheng Chen}
\authornotemark[1]
\email{zgchen@amazon.com}
\orcid{0000-0003-4406-2193}
\affiliation{%
  \institution{Amazon Alexa AI}
  \city{Seattle}
  \state{WA}
  \country{USA}
}

\author{Ziyan Jiang}
\authornotemark[1]
\email{ziyjiang@amazon.com}
\orcid{1234-5678-9012}
\affiliation{%
  \institution{Amazon Alexa AI}
  \city{Seattle}
  \state{WA}
  \country{USA}
}

\author{Eunah Cho}
\email{eunahch@amazon.com}
\affiliation{%
  \institution{Amazon Alexa AI}
  \city{Seattle}
  \state{WA}
  \country{USA}
}

\author{Xiaojiang Huang}
\email{xjhuang@amazon.com}
\affiliation{%
  \institution{Amazon Alexa AI}
  \city{Seattle}
  \state{WA}
  \country{USA}
}

\author{Yanbin Lu}
\email{luyanbin@amazon.com}
\affiliation{%
  \institution{Amazon Alexa AI}
  \city{Seattle}
  \state{WA}
  \country{USA}
}

\renewcommand{\shortauthors}{Trovato and Tobin, et al.}

\begin{abstract}
Large language models (LLMs) have recently received significant attention for their exceptional capabilities. Despite extensive efforts in developing general-purpose LLMs that can be utilized in various natural language processing (NLP) tasks, there has been less research exploring their potential in recommender systems. In this paper, we propose a novel framework, named \textbf{PALR} (\underline{\textbf{P}}ersonalization \underline{\textbf{A}}ware \underline{\textbf{L}}LMs for \underline{\textbf{R}}ecommendation), aimed at integrating user history behaviors (such as clicks, purchases, ratings, etc.) with LLMs to generate user preferred items. Specifically, we first use user/item interactions as guidance for candidate retrieval, and then adopt an LLM-based ranking model to generate recommended items. Unlike existing approaches that typically adopt general-purpose LLMs for zero/few-shot recommendation testing or training on small-sized language models (with less than 1 billion parameters), which cannot fully elicit LLMs' reasoning abilities and leverage rich item side parametric knowledge, we fine-tune an LLM of 7 billion parameters for the ranking purpose. This model takes retrieval candidates in natural language format as input, with instructions explicitly asking to select results from input candidates during inference. Our experimental results demonstrate that our solution outperforms state-of-the-art models on various sequential recommendation tasks.
\end{abstract}



\received{20 February 2007}
\received[revised]{12 March 2009}
\received[accepted]{5 June 2009}

\maketitle

\section{Introduction}

A recommender system is a type of information filtering system that is designed to predict and recommend items or products. These systems are widely used in e-commerce, online advertising, social media, and entertainment industries. In recent years, the emergence of LLMs, such as Bert\cite{Devlin2019BERTPO}, GPT-3\cite{Brown2020LanguageMA}, FLAN-T5\cite{chung2022scaling}, has led to significant breakthroughs in NLP research. Inspired by these advancements, researchers have begun exploring the potential of using LLMs in recommendation systems\cite{Sun2019BERT4RecSR, cui2022m6, li2023gpt4rec,liu2023chatgpt}. 

Compared with traditional recommendation modeling techniques \cite{Shani2002AnMR,Rendle2010FactorizingPM,Hidasi2015SessionbasedRW} and more recent sequential modeling\cite{kang2018self, Sun2019BERT4RecSR, Tan2021DynamicMB} and graph modeling\cite{Xiao2019HierarchicalNV, he2020lightgcn} techniques, LLMs offer several distinct advantages. Firstly, LLMs inherently support inductive learning and negate the need for pre-trained embeddings for each item. Instead, each item can be represented as a piece of text. This feature is particularly crucial in an industry setting where new items are continually emerging. Secondly, LLMs allow for easy integration of various signals, such as metadata, context, and multi-modal signals, into the recommendation process by incorporating them into the model's prompt. Thirdly, LLMs can transfer knowledge acquired from one domain to another, providing a significant advantage in cold-start scenarios where user behavior data is limited. Finally, LLMs possess vast knowledge and superior reasoning capabilities given their extensive pre-training, along with the ability to generate natural language outputs; therefore they can make recommendations with sensible and human-readable explanations, enhancing user trust and engagement.

However, directly leveraging parametric knowledge saved in general-purpose LLMs\cite{touvron2023llama,chung2022scaling} to generate recommended items is challenging\cite{liu2023chatgpt}. Firstly, there may be knowledge gaps between LLMs and items that need to be recommended. For instance, some newly released shopping items may not be included in the LLM's parametric knowledge. Secondly, LLMs are prone to generating incomplete and hallucinatory results, which need an extra knowledge grounding step to eliminate unfavorable results. Thirdly, LLMs have limitations regarding input token length and efficiency. If the items pool is extensive, it is impractical to provide all items as natural language input. As a result, recent research is pruned to treat LLMs as summarization and reasoning engines instead of a knowledge base in recommendation scenarios. 

In this paper, we present PALR, which is a general framework for personalized recommendation tasks that combines user behaviors with LLM. Given the challenges mentioned above, we break down the task into several stages. Initially, we use an LLM and user behavior as input to generate user profile keywords. Then, we employ a retrieval module to pre-filter some candidates from the items pool based on the user profile. Importantly, our framework is not reliant on any specific retrieval algorithms. Finally, we use LLM to provide recommendations from those candidates based on user history behaviors. To better adapt these general-purpose LLMs to fit the recommendation scenarios, we convert user behavior data into natural language prompts and fine-tune a LLaMa\cite{touvron2023llama} 7B model. Our goal is to teach the model to learn the co-occurrence of user engaged item patterns. This approach enables us to incorporate user behavior data into the LLM' reasoning process and better generalize to new users and unseen items. In summary, our contributions are: 
\begin{enumerate}
\item We propose PALR, a flexible personalized recommendation framework, which incorporating user behaviors with LLMs for recommended items generation. 
\item We break down recommendation task into user profile generation, candidates retrieval and items ranking three sub-tasks, and tune instruction prompts to better elicit LLMs reasoning ability.  
\item We fine-tune a recommendation oriented LLM based on LLaMa 7B. Evaluation under PALR framework on two public datasets demonstrate its competitive performance against state-of-the-art methods. 
\item We experimented with two datasets, MovieLens-1M\cite{harper2015movielens}, and Amazon Beauty\cite{ni2019justifying} and demonstrated the strong potential of an LLM for recommendation in comparison to SOTA.
\end{enumerate}

\section{Methodology}

\begin{figure}
    \centering
    \includegraphics[width=\columnwidth]{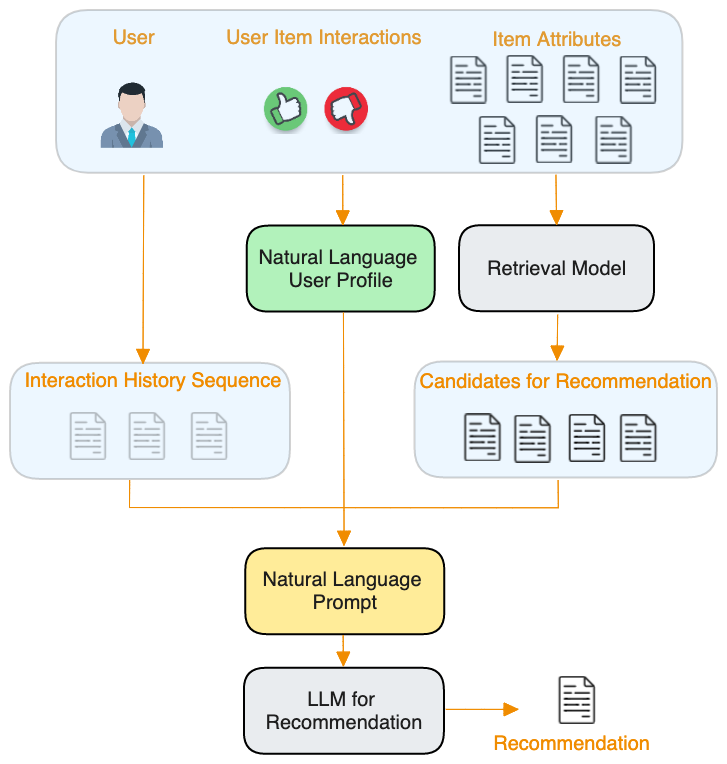}
    \caption{Here is an overview of our proposed PALR architecture. The "Natural Language Prompt" for the "LLM for recommendation" comprises three components: the "Interaction History Sequence," the "Natural Language User Profile," and the "Candidates for Recommendation". The "Interaction History Sequence" is created by simply concatenating the items that the user has interacted with. The "Natural Language User Profile" is a high-level summarization of the user's preferences, generated using an LLM based on user-item interactions, item attributes, or even user information if possible. The "Candidates for Recommendation" are the output of a retrieval model, and in our design, we have the flexibility to use any retrieval model for this purpose. We have included an example in Figure \ref{fig:prompt example}.}
    \label{fig:prompt example}
\end{figure}

\begin{figure*}
    \centering
    \includegraphics[width=\textwidth]{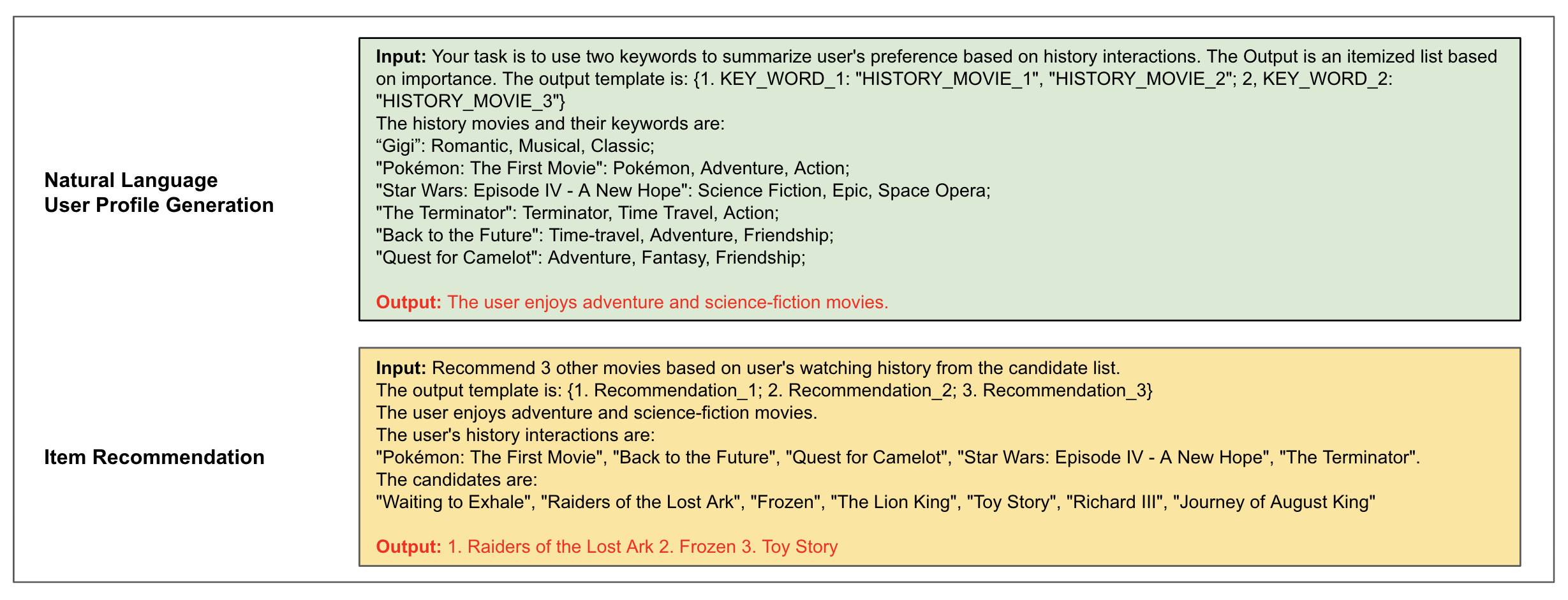}
    \caption{Here is an example for movie recommendation. (Above) The "Natural Language User Profile" leverages an LLM to summarize the user's preferences by taking into account the movies the user has watched and the movie keywords. (Below) The "LLM for Recommendation" takes the "Natural Language Prompt" as its input, which is composed of three parts: the "Interaction History Sequence," the "Natural Language User Profile," and the "Candidates for Recommendation."}
    \label{fig:prompt example}
\end{figure*}

\subsection{PALR Framework}
Our proposed method, PALR (Personalization Aware LLM for Recommendation), is illustrated in Figure
1. It utilizes a multi-step approach to harness the potential of LLMs for recommendation.
\begin{itemize}
\item \textbf{Natural Language user profile generation}. When a user interacts with a large number of items and exhibits mixed affinities, it can be challenging for the model to provide accurate recommendations based solely on user behaviors. In such situations, having a high-level summarization of the user's preferences can be beneficial. An LLM can be leveraged to generate a summary of a user's preferences. For example, by analyzing a user's music and TV viewing history, we can generate a summary of their preferences such as "pop music" and "fantasy movies."
\item \textbf{Candidates retrieval}. To address the issues of hallucination and incompleteness in the generated results, a retrieval module is utilized to ground knowledge and filter out results that do not relevant to the task at hand, resulting in a much smaller candidate pool to feed into the LLM for further processing. This framework can accommodate various retrieval models, such as a sequential recommendation model trained on user behaviors, which can serve this purpose effectively.
\item \textbf{Item recommendation}. By combining the interaction history, natural language user profile and retrieved candidates, we can create a natural language prompt that can be fed into the LLM for recommendation. The model will utilize its reasoning ability to select the items from the candidates pool that align best with user profile.
\end{itemize}

The steps of "user profile generation" and "item recommendation" require dedicated prompt design to effectively leverage the reasoning ability of LLMs. An example of related prompt design in the movie recommendation task is shown in Figure 2.

\subsection{Fine-Tuning}

Through our investigation, we find fine-tuning is necessary to make the model 1) obtain reasonably strong performance, and 2) recognize the retrieval layer and performs the retrieval as expected. We employ instruction-based fine-tuning, a technique proven effective in recent LLM development \cite{alpaca, Wei2021FinetunedLM, Wang2022SelfInstructAL}. 

We have created two types of instruction tasks called "Recommend" and "Recommend\_Retrieval". The "Recommend" task involves a list of items that the user has interacted with in the past (with a maximum limit of 20 items), and the objective of the model is to generate a list of "future" items that the user may interact with. Here's an example of such an instruction for the Movielens dataset. We refer to a model fine-tuned by this instruction as $PALR_{v1}$.
\begin{center} 
    \fbox{%
        \parbox{0.8\columnwidth}{
            \textbf{Task Instruction:} Recommend 10 other movies based on user's watching history.
            
            \textbf{Input:} User watched movies "Pink Floyd - The Wall", "Canadian Bacon", "G.I. Jane", ..., "Down by Law".

            \textbf{Output:} "Almost Famous", "Full Metal Jacket", ...
        }
    }
\end{center}

The "Recommend\_Retrieval" task asks the model to retrieve the target "future" items from a list of candidate items. The candidate list contains all target items, plus a few negative items similar to the target items (e.g. movies with the same genres, co-watched by many users). The following are two examples of such instruction used in our fine-tuning for the Movielens dataset and the Amazon Beauty dataset. For the Amazon beauty dataset, we include item ID for evaluation.  We refer to a model fine-tuned with both "Recommend" and "Recommend\_Retrieval" instruction as $PALR_{v2}$.

\begin{center} 
    \fbox{%
        \parbox{0.8\columnwidth}{
            \textbf{Task Instruction}: Recommend 10 other movies based on user's watching history from the candidate list.
            
            \textbf{Input:} User watched movies "Pink Floyd - The Wall", "Canadian Bacon", "G.I. Jane", ..., "Down by Law". The candidates are “The Blues Brothers”, "Platoon", ..., "Almost Famous".

            \textbf{Output:} "Almost Famous", "Full Metal Jacket", ...
        }
    }
\end{center}

\begin{center} 
    \fbox{%
        \parbox{0.8\columnwidth}{
            \textbf{Task Instruction}: Recommend 10 other items based on user's history from the candidate list.
            
             \textbf{Input:} User has purchased the following products "B000142FVW (Opi Nail Lacquer, Not So Bora Pink, 0.5 Fluid Ounce)", ..., "B001PPMLS8 (MoroccanOil Treatment Light 3.4 Oz)". The candidates are "B000SQRMOS (FHI Heat Hot Sauce (50ml), 1.7 fluid ounces bottle)", ..., "B000G1MT2U (Mixed Chicks Leave-In Conditioner)".

            \textbf{Output:} "B000G1MT2U (Mixed Chicks Leave-In Conditioner)", "B001AO0WCG (Moroccan Oil Hair Treatment 3.4 Oz Bottle with Blue Box)", ...
        }
    }
\end{center}

It is worth noting that the fine-tuning is retrieval-layer agnostic. Despite our objective being to train the model to select from a list of candidates, the construction of this list for fine-tuning is not bound to the retrieval layer in our framework.

Furthermore, we have found that the fine-tuning process is enhanced by a couple of techniques: 1) enriching shorter lists in the datasets with items from the user's 3-hop affinity; 2) randomly swapping items between the instruction and the generation label.

Last but not the least, we only fine-tune on 20\% of users. We intend to demo the strong inductive learning capabilities of LLMs. This is not possible for item-embedding based models such as \cite{Kang2018SelfAttentiveSR,Sun2019BERT4RecSR}, which must be trained on the full data to function effectively.

\section{Experiments}

\subsection{Experiments Settings}
\subsubsection{Datasets}
The two public datasets are collected from the real-world platforms and have been widely used for sequential recommendation. \textit{Amazon Beauty}\footnote{\url{https://cseweb.ucsd.edu/~jmcauley/datasets/amazon_v2/}} is one category of Amazon review datasets, which contains a collection of user-item interactions on Amazon spanning from May 1996 to July 2014. \textit{Movielens-1M}\footnote{\url{http://files.grouplens.org/datasets/movielens/ml-1m.zip}}
is a common benchmark dataset that includes one million movie ratings.

For dataset preprocessing, we follow the common practice\cite{Kang2018SelfAttentiveSR,Sun2019BERT4RecSR}. We convert all numeric ratings or presence of a review to “1” and others to “0”. Then, for each user, we discard duplicated interactions and then sort their historical items by the interacted time step chronologically to obtain the user interacted sequence. It is worth mentioning that to guarantee each user/item with enough interactions, we follow the preprocessing procedure in\cite{He2017NeuralCF, Kang2018SelfAttentiveSR}, which only keeps the “5-core” datasets. We discard users and items with fewer than 5 interaction records iteratively. The statistics of these datasets are reported in Table \ref{tab:stats}.

\begin{table}
\centering
\begin{tabular}{lccc}
\hline
\textbf{Dataset} & \textbf{\# Users} & \textbf{\# Items} & \textbf{\# Interactions} \\
\hline
Beauty & 22,363 & 12,101 & 198,502 \\
Movielens-1M & 6,040 &3,416 & 999,611  \\
\hline
\end{tabular}
\caption{Statistics of datasets after preprocessing.}
\label{tab:stats}
\end{table}

\subsubsection{Evaluation}
We adopt the leave-one-out strategy to evaluate the performance of each method, which is widely employed in many related works. For each user, we hold out the
last interacted item as the test data and utilize the item just before the last as the validation data. The remaining items are used for training. We evaluate each method on the whole item set without sampling as suggested in previous study \cite{Krichene2020OnSM}. We employ Hit Ratio (HR) and Normalized Discounted Cumulative Gain (NDCG) to evaluate the performance. HR focuses on the presence
of the positive item, while NDCG further takes the rank position information into account.

\subsubsection{Baselines}
To verify the effectiveness of our method, we compare it with the following representative baselines.
\begin{itemize}
\item \textbf{BPR-MF}\cite{Rendle2009BPRBP}. It utilizes matrix factorization to model users and items with the pairwise Bayesian Personalized Ranking (BPR) loss.
\item \textbf{NCF}\cite{He2017NeuralCF}. It employs a neural network architecture to model non-sequential user-item interactions instead of the inner product used by matrix factorization.
\item \textbf{GRU4Rec}\cite{Hidasi2015SessionbasedRW}. It utilizes GRU to model the sequential behavior of users for recommendation.
\item \textbf{Caser}\cite{Tang2018PersonalizedTS}. It devises horizontal and vertical CNN to exploit user’s recent sub-sequence behaviors for recommendation.
\item \textbf{SASRec}\cite{Kang2018SelfAttentiveSR}. It models user sequences through self-attention modules to capture users’ dynamic interests. and it is a competitive benchmark in sequential recommendation.

\end{itemize}



\subsection{Overall Performance Comparison}
Table \ref{tab:seq_req} summarizes the best results of all models on two benchmark datasets. As shown in Table \ref{tab:seq_req}, our $PALR_{v2}$ outperforms multiple baselines by a large margin on two datasets.
A comparison between  $PALR_{v1}$ and  $PALR_{v2}$ reveals the crucial role of candidates retrieval in improving performance. As we mentioned before, our framework does not depend on any particular retrieval algorithms. Ideally, PALR can function as an effective ranking model in conjunction with various retrieval methods.In this paper, we utilize SASRec as our retrieval layer and consider its top 50 recommendations. By comparing $PALR_{v2}$ and SASRec, it's obvious that the top10 recommendations re-ranked by our PALR are superior to the original recommendations provided by SASRec. We also evaluate our framework using different recommendation algorithms, including BERT4Rec and LightGCN, and observe a similar trend.

By conducting various experiments, we are able to gain a deeper understanding of the significance of fine-tuning. We could observe $PALR_{v1}$ has shown some ability to connect historical interacted items with possible future interacted items. Prior to fine-tuning, the model tends to only recommend popular movies in movie recommendation tasks. However, $PALR_{v1}$ isn't able to retrieve the target item from a list of candidate items. We have tried to use $PALR_{v1}$ for retrieval and observe that it could only randomly select from the candidates. The performance from $PALR_{v2}$ has demonstrated the effectiveness of incorporating an additional instruction during the fine-tuning stage.


\begin{table}
\centering
\begin{tabular}{lccccc}
\hline
\hline
Dataset & Model & HR@10 & NDCG@10  \\
\hline
            \multirow{7}{*}{Beauty} & \multicolumn{1}{l}{BPR-MF} & \multicolumn{1}{c}{0.0299} & \multicolumn{1}{c}{0.0122} \\\cline{2-4}
                                 & \multicolumn{1}{l}{NCF} & \multicolumn{1}{c}{0.0293} & \multicolumn{1}{c}{0.0130} \\\cline{2-4}
                                 & \multicolumn{1}{l}{GRU4Rec} & \multicolumn{1}{c}{0.0194} & \multicolumn{1}{c}{0.0091}    \\\cline{2-4}
                                 & \multicolumn{1}{l}{Caser} & \multicolumn{1}{c}{0.0282} & \multicolumn{1}{c}{0.0136}    \\\cline{2-4}
                                 & \multicolumn{1}{l}{SASRec} & \multicolumn{1}{c}{0.0617} & \multicolumn{1}{c}{0.0283}    \\\cline{2-4}
                                 & \multicolumn{1}{l}{$PALR_{v1}$} & \multicolumn{1}{c}{0.0181} & \multicolumn{1}{c}{0.0101}    \\\cline{2-4}
                                 & \multicolumn{1}{l}{$PALR_{v2}$} & \multicolumn{1}{c}{\textbf{0.0721}} & \multicolumn{1}{c}{\textbf{0.0446}}    \\\hline     
            \multirow{7}{*}{ML-1M} & \multicolumn{1}{l}{BPR-MF} & \multicolumn{1}{c}{0.0354} & \multicolumn{1}{c}{0.0158}                         \\\cline{2-4}
                                 & \multicolumn{1}{l}{NCF} & \multicolumn{1}{c}{0.0313} & \multicolumn{1}{c}{0.0143} \\\cline{2-4}
                                 & \multicolumn{1}{l}{GRU4Rec} & \multicolumn{1}{c}{0.1017} & \multicolumn{1}{c}{0.0468}    \\\cline{2-4}
                                 & \multicolumn{1}{l}{Caser} & \multicolumn{1}{c}{0.1338} & \multicolumn{1}{c}{0.0614}    \\\cline{2-4}
                                 & \multicolumn{1}{l}{SASRec} & \multicolumn{1}{c}{0.1978} & \multicolumn{1}{c}{0.1192}    \\\cline{2-4}
                                 & \multicolumn{1}{l}{$PALR_{v1}$} & \multicolumn{1}{c}{0.1216} & \multicolumn{1}{c}{0.0569}    \\\cline{2-4}
                                 & \multicolumn{1}{l}{$PALR_{v2}$} & \multicolumn{1}{c}{\textbf{0.2110}} & \multicolumn{1}{c}{\textbf{0.1276}}    \\\hline    
\hline
\end{tabular}
\caption{Experimental results on the two datasets. The best results are in boldface.}
\label{tab:seq_req}
\end{table}

\section{Conclusion}
The paper introduces PALR, a novel generative framework for producing personalized recommendations, which utilizes a multi-step paradigm to better utilize the knowledge in LLMs' parameters and reasoning abilities for sequential recommendation tasks. Additionally, the paper discusses the recent advances in LLMs and how they can be leveraged for recommendation tasks. Besides its competitive experiment results mentioned in the paper, LLMs has some other unique benefits in the recommendation task. The first advantage of using LLMs in recommendation tasks is the ease with which external knowledge from different sources can be incorporated into the framework. The second advantage is that LLMs offer an easier pathway to more complex recommendation scenarios, including explainable recommendations and conversational recommendations. Moving forward, our research will focus on further leveraging LLMs in recommendation tasks while ensuring a balance between their powerful capabilities and latency. As LLMs can be computationally intensive, we will explore ways to optimize their performance and reduce latency without sacrificing accuracy or personalization.

\section*{Acknowledgement}
We thank the LLaMA team for giving us access to their models.

\bibliographystyle{ACM-Reference-Format}
\bibliography{custom}


\begin{thebibliography}{25}


\ifx \showCODEN    \undefined \def \showCODEN     #1{\unskip}     \fi
\ifx \showDOI      \undefined \def \showDOI       #1{#1}\fi
\ifx \showISBNx    \undefined \def \showISBNx     #1{\unskip}     \fi
\ifx \showISBNxiii \undefined \def \showISBNxiii  #1{\unskip}     \fi
\ifx \showISSN     \undefined \def \showISSN      #1{\unskip}     \fi
\ifx \showLCCN     \undefined \def \showLCCN      #1{\unskip}     \fi
\ifx \shownote     \undefined \def \shownote      #1{#1}          \fi
\ifx \showarticletitle \undefined \def \showarticletitle #1{#1}   \fi
\ifx \showURL      \undefined \def \showURL       {\relax}        \fi
\providecommand\bibfield[2]{#2}
\providecommand\bibinfo[2]{#2}
\providecommand\natexlab[1]{#1}
\providecommand\showeprint[2][]{arXiv:#2}

\bibitem[Brown et~al\mbox{.}(2020)]%
        {Brown2020LanguageMA}
\bibfield{author}{\bibinfo{person}{Tom~B. Brown}, \bibinfo{person}{Benjamin
  Mann}, \bibinfo{person}{Nick Ryder}, \bibinfo{person}{Melanie Subbiah},
  \bibinfo{person}{Jared Kaplan}, \bibinfo{person}{Prafulla Dhariwal},
  \bibinfo{person}{Arvind Neelakantan}, \bibinfo{person}{Pranav Shyam},
  \bibinfo{person}{Girish Sastry}, \bibinfo{person}{Amanda Askell},
  \bibinfo{person}{Sandhini Agarwal}, \bibinfo{person}{Ariel Herbert-Voss},
  \bibinfo{person}{Gretchen Krueger}, \bibinfo{person}{T.~J. Henighan},
  \bibinfo{person}{Rewon Child}, \bibinfo{person}{Aditya Ramesh},
  \bibinfo{person}{Daniel~M. Ziegler}, \bibinfo{person}{Jeff Wu},
  \bibinfo{person}{Clemens Winter}, \bibinfo{person}{Christopher Hesse},
  \bibinfo{person}{Mark Chen}, \bibinfo{person}{Eric Sigler},
  \bibinfo{person}{Mateusz Litwin}, \bibinfo{person}{Scott Gray},
  \bibinfo{person}{Benjamin Chess}, \bibinfo{person}{Jack Clark},
  \bibinfo{person}{Christopher Berner}, \bibinfo{person}{Sam McCandlish},
  \bibinfo{person}{Alec Radford}, \bibinfo{person}{Ilya Sutskever}, {and}
  \bibinfo{person}{Dario Amodei}.} \bibinfo{year}{2020}\natexlab{}.
\newblock \showarticletitle{Language Models are Few-Shot Learners}.
\newblock \bibinfo{journal}{\emph{ArXiv}}  \bibinfo{volume}{abs/2005.14165}
  (\bibinfo{year}{2020}).
\newblock


\bibitem[Chung et~al\mbox{.}(2022)]%
        {chung2022scaling}
\bibfield{author}{\bibinfo{person}{Hyung~Won Chung}, \bibinfo{person}{Le Hou},
  \bibinfo{person}{Shayne Longpre}, \bibinfo{person}{Barret Zoph},
  \bibinfo{person}{Yi Tay}, \bibinfo{person}{William Fedus},
  \bibinfo{person}{Eric Li}, \bibinfo{person}{Xuezhi Wang},
  \bibinfo{person}{Mostafa Dehghani}, \bibinfo{person}{Siddhartha Brahma},
  {et~al\mbox{.}}} \bibinfo{year}{2022}\natexlab{}.
\newblock \showarticletitle{Scaling instruction-finetuned language models}.
\newblock \bibinfo{journal}{\emph{arXiv preprint arXiv:2210.11416}}
  (\bibinfo{year}{2022}).
\newblock


\bibitem[Cui et~al\mbox{.}(2022)]%
        {cui2022m6}
\bibfield{author}{\bibinfo{person}{Zeyu Cui}, \bibinfo{person}{Jianxin Ma},
  \bibinfo{person}{Chang Zhou}, \bibinfo{person}{Jingren Zhou}, {and}
  \bibinfo{person}{Hongxia Yang}.} \bibinfo{year}{2022}\natexlab{}.
\newblock \showarticletitle{M6-Rec: Generative Pretrained Language Models are
  Open-Ended Recommender Systems}.
\newblock \bibinfo{journal}{\emph{arXiv preprint arXiv:2205.08084}}
  (\bibinfo{year}{2022}).
\newblock


\bibitem[Devlin et~al\mbox{.}(2019)]%
        {Devlin2019BERTPO}
\bibfield{author}{\bibinfo{person}{Jacob Devlin}, \bibinfo{person}{Ming-Wei
  Chang}, \bibinfo{person}{Kenton Lee}, {and} \bibinfo{person}{Kristina
  Toutanova}.} \bibinfo{year}{2019}\natexlab{}.
\newblock \showarticletitle{BERT: Pre-training of Deep Bidirectional
  Transformers for Language Understanding}.
\newblock \bibinfo{journal}{\emph{ArXiv}}  \bibinfo{volume}{abs/1810.04805}
  (\bibinfo{year}{2019}).
\newblock


\bibitem[Harper and Konstan(2015)]%
        {harper2015movielens}
\bibfield{author}{\bibinfo{person}{F~Maxwell Harper} {and}
  \bibinfo{person}{Joseph~A Konstan}.} \bibinfo{year}{2015}\natexlab{}.
\newblock \showarticletitle{The movielens datasets: History and context}.
\newblock \bibinfo{journal}{\emph{Acm transactions on interactive intelligent
  systems (tiis)}} \bibinfo{volume}{5}, \bibinfo{number}{4}
  (\bibinfo{year}{2015}), \bibinfo{pages}{1--19}.
\newblock


\bibitem[He et~al\mbox{.}(2020)]%
        {he2020lightgcn}
\bibfield{author}{\bibinfo{person}{Xiangnan He}, \bibinfo{person}{Kuan Deng},
  \bibinfo{person}{Xiang Wang}, \bibinfo{person}{Yan Li},
  \bibinfo{person}{Yongdong Zhang}, {and} \bibinfo{person}{Meng Wang}.}
  \bibinfo{year}{2020}\natexlab{}.
\newblock \showarticletitle{Lightgcn: Simplifying and powering graph
  convolution network for recommendation}. In
  \bibinfo{booktitle}{\emph{Proceedings of the 43rd International ACM SIGIR
  conference on research and development in Information Retrieval}}.
  \bibinfo{pages}{639--648}.
\newblock


\bibitem[He et~al\mbox{.}(2017)]%
        {He2017NeuralCF}
\bibfield{author}{\bibinfo{person}{Xiangnan He}, \bibinfo{person}{Lizi Liao},
  \bibinfo{person}{Hanwang Zhang}, \bibinfo{person}{Liqiang Nie},
  \bibinfo{person}{Xia Hu}, {and} \bibinfo{person}{Tat-Seng Chua}.}
  \bibinfo{year}{2017}\natexlab{}.
\newblock \showarticletitle{Neural Collaborative Filtering}.
\newblock \bibinfo{journal}{\emph{Proceedings of the 26th International
  Conference on World Wide Web}} (\bibinfo{year}{2017}).
\newblock


\bibitem[Hidasi et~al\mbox{.}(2015)]%
        {Hidasi2015SessionbasedRW}
\bibfield{author}{\bibinfo{person}{Bal{\'a}zs Hidasi},
  \bibinfo{person}{Alexandros Karatzoglou}, \bibinfo{person}{Linas Baltrunas},
  {and} \bibinfo{person}{Domonkos Tikk}.} \bibinfo{year}{2015}\natexlab{}.
\newblock \showarticletitle{Session-based Recommendations with Recurrent Neural
  Networks}.
\newblock \bibinfo{journal}{\emph{CoRR}}  \bibinfo{volume}{abs/1511.06939}
  (\bibinfo{year}{2015}).
\newblock


\bibitem[Kang and McAuley(2018a)]%
        {kang2018self}
\bibfield{author}{\bibinfo{person}{Wang-Cheng Kang} {and}
  \bibinfo{person}{Julian McAuley}.} \bibinfo{year}{2018}\natexlab{a}.
\newblock \showarticletitle{Self-attentive sequential recommendation}. In
  \bibinfo{booktitle}{\emph{2018 IEEE international conference on data mining
  (ICDM)}}. IEEE, \bibinfo{pages}{197--206}.
\newblock


\bibitem[Kang and McAuley(2018b)]%
        {Kang2018SelfAttentiveSR}
\bibfield{author}{\bibinfo{person}{Wang-Cheng Kang} {and}
  \bibinfo{person}{Julian McAuley}.} \bibinfo{year}{2018}\natexlab{b}.
\newblock \showarticletitle{Self-Attentive Sequential Recommendation}.
\newblock \bibinfo{journal}{\emph{2018 IEEE International Conference on Data
  Mining (ICDM)}} (\bibinfo{year}{2018}), \bibinfo{pages}{197--206}.
\newblock


\bibitem[Krichene and Rendle(2020)]%
        {Krichene2020OnSM}
\bibfield{author}{\bibinfo{person}{Walid Krichene} {and}
  \bibinfo{person}{Steffen Rendle}.} \bibinfo{year}{2020}\natexlab{}.
\newblock \showarticletitle{On Sampled Metrics for Item Recommendation}.
\newblock \bibinfo{journal}{\emph{Proceedings of the 26th ACM SIGKDD
  International Conference on Knowledge Discovery \& Data Mining}}
  (\bibinfo{year}{2020}).
\newblock


\bibitem[Li et~al\mbox{.}(2023)]%
        {li2023gpt4rec}
\bibfield{author}{\bibinfo{person}{Jinming Li}, \bibinfo{person}{Wentao Zhang},
  \bibinfo{person}{Tian Wang}, \bibinfo{person}{Guanglei Xiong},
  \bibinfo{person}{Alan Lu}, {and} \bibinfo{person}{Gerard Medioni}.}
  \bibinfo{year}{2023}\natexlab{}.
\newblock \showarticletitle{GPT4Rec: A Generative Framework for Personalized
  Recommendation and User Interests Interpretation}.
\newblock \bibinfo{journal}{\emph{arXiv preprint arXiv:2304.03879}}
  (\bibinfo{year}{2023}).
\newblock


\bibitem[Liu et~al\mbox{.}(2023)]%
        {liu2023chatgpt}
\bibfield{author}{\bibinfo{person}{Junling Liu}, \bibinfo{person}{Chao Liu},
  \bibinfo{person}{Renjie Lv}, \bibinfo{person}{Kang Zhou}, {and}
  \bibinfo{person}{Yan Zhang}.} \bibinfo{year}{2023}\natexlab{}.
\newblock \showarticletitle{Is ChatGPT a Good Recommender? A Preliminary
  Study}.
\newblock \bibinfo{journal}{\emph{arXiv preprint arXiv:2304.10149}}
  (\bibinfo{year}{2023}).
\newblock


\bibitem[Ni et~al\mbox{.}(2019)]%
        {ni2019justifying}
\bibfield{author}{\bibinfo{person}{Jianmo Ni}, \bibinfo{person}{Jiacheng Li},
  {and} \bibinfo{person}{Julian McAuley}.} \bibinfo{year}{2019}\natexlab{}.
\newblock \showarticletitle{Justifying recommendations using distantly-labeled
  reviews and fine-grained aspects}. In \bibinfo{booktitle}{\emph{Proceedings
  of the 2019 conference on empirical methods in natural language processing
  and the 9th international joint conference on natural language processing
  (EMNLP-IJCNLP)}}. \bibinfo{pages}{188--197}.
\newblock


\bibitem[Rendle et~al\mbox{.}(2009)]%
        {Rendle2009BPRBP}
\bibfield{author}{\bibinfo{person}{Steffen Rendle}, \bibinfo{person}{Christoph
  Freudenthaler}, \bibinfo{person}{Zeno Gantner}, {and} \bibinfo{person}{Lars
  Schmidt-Thieme}.} \bibinfo{year}{2009}\natexlab{}.
\newblock \showarticletitle{BPR: Bayesian Personalized Ranking from Implicit
  Feedback}.
\newblock \bibinfo{journal}{\emph{ArXiv}}  \bibinfo{volume}{abs/1205.2618}
  (\bibinfo{year}{2009}).
\newblock


\bibitem[Rendle et~al\mbox{.}(2010)]%
        {Rendle2010FactorizingPM}
\bibfield{author}{\bibinfo{person}{Steffen Rendle}, \bibinfo{person}{Christoph
  Freudenthaler}, {and} \bibinfo{person}{Lars Schmidt-Thieme}.}
  \bibinfo{year}{2010}\natexlab{}.
\newblock \showarticletitle{Factorizing personalized Markov chains for
  next-basket recommendation}. In \bibinfo{booktitle}{\emph{The Web
  Conference}}.
\newblock


\bibitem[Shani et~al\mbox{.}(2002)]%
        {Shani2002AnMR}
\bibfield{author}{\bibinfo{person}{Guy Shani}, \bibinfo{person}{David~E.
  Heckerman}, {and} \bibinfo{person}{Ronen~I. Brafman}.}
  \bibinfo{year}{2002}\natexlab{}.
\newblock \showarticletitle{An MDP-Based Recommender System}.
\newblock \bibinfo{journal}{\emph{ArXiv}}  \bibinfo{volume}{abs/1301.0600}
  (\bibinfo{year}{2002}).
\newblock


\bibitem[Sun et~al\mbox{.}(2019)]%
        {Sun2019BERT4RecSR}
\bibfield{author}{\bibinfo{person}{Fei Sun}, \bibinfo{person}{Jun Liu},
  \bibinfo{person}{Jian Wu}, \bibinfo{person}{Changhua Pei},
  \bibinfo{person}{Xiao Lin}, \bibinfo{person}{Wenwu Ou}, {and}
  \bibinfo{person}{Peng Jiang}.} \bibinfo{year}{2019}\natexlab{}.
\newblock \showarticletitle{BERT4Rec: Sequential Recommendation with
  Bidirectional Encoder Representations from Transformer}.
\newblock \bibinfo{journal}{\emph{Proceedings of the 28th ACM International
  Conference on Information and Knowledge Management}} (\bibinfo{year}{2019}).
\newblock


\bibitem[Tan et~al\mbox{.}(2021)]%
        {Tan2021DynamicMB}
\bibfield{author}{\bibinfo{person}{Qiaoyu Tan}, \bibinfo{person}{Jianwei
  Zhang}, \bibinfo{person}{Ninghao Liu}, \bibinfo{person}{Xiao Huang},
  \bibinfo{person}{Hongxia Yang}, \bibinfo{person}{Jingren Zhou}, {and}
  \bibinfo{person}{Xia Hu}.} \bibinfo{year}{2021}\natexlab{}.
\newblock \showarticletitle{Dynamic Memory based Attention Network for
  Sequential Recommendation}. In \bibinfo{booktitle}{\emph{AAAI Conference on
  Artificial Intelligence}}.
\newblock


\bibitem[Tang and Wang(2018)]%
        {Tang2018PersonalizedTS}
\bibfield{author}{\bibinfo{person}{Jiaxi Tang} {and} \bibinfo{person}{Ke
  Wang}.} \bibinfo{year}{2018}\natexlab{}.
\newblock \showarticletitle{Personalized Top-N Sequential Recommendation via
  Convolutional Sequence Embedding}.
\newblock \bibinfo{journal}{\emph{Proceedings of the Eleventh ACM International
  Conference on Web Search and Data Mining}} (\bibinfo{year}{2018}).
\newblock


\bibitem[Taori et~al\mbox{.}(2023)]%
        {alpaca}
\bibfield{author}{\bibinfo{person}{Rohan Taori}, \bibinfo{person}{Ishaan
  Gulrajani}, \bibinfo{person}{Tianyi Zhang}, \bibinfo{person}{Yann Dubois},
  \bibinfo{person}{Xuechen Li}, \bibinfo{person}{Carlos Guestrin},
  \bibinfo{person}{Percy Liang}, {and} \bibinfo{person}{Tatsunori~B.
  Hashimoto}.} \bibinfo{year}{2023}\natexlab{}.
\newblock \bibinfo{title}{Stanford Alpaca: An Instruction-following LLaMA
  model}.
\newblock
  \bibinfo{howpublished}{\url{https://github.com/tatsu-lab/stanford_alpaca}}.
\newblock


\bibitem[Touvron et~al\mbox{.}(2023)]%
        {touvron2023llama}
\bibfield{author}{\bibinfo{person}{Hugo Touvron}, \bibinfo{person}{Thibaut
  Lavril}, \bibinfo{person}{Gautier Izacard}, \bibinfo{person}{Xavier
  Martinet}, \bibinfo{person}{Marie-Anne Lachaux},
  \bibinfo{person}{Timoth{\'e}e Lacroix}, \bibinfo{person}{Baptiste
  Rozi{\`e}re}, \bibinfo{person}{Naman Goyal}, \bibinfo{person}{Eric Hambro},
  \bibinfo{person}{Faisal Azhar}, {et~al\mbox{.}}}
  \bibinfo{year}{2023}\natexlab{}.
\newblock \showarticletitle{Llama: Open and efficient foundation language
  models}.
\newblock \bibinfo{journal}{\emph{arXiv preprint arXiv:2302.13971}}
  (\bibinfo{year}{2023}).
\newblock


\bibitem[Wang et~al\mbox{.}(2022)]%
        {Wang2022SelfInstructAL}
\bibfield{author}{\bibinfo{person}{Yizhong Wang}, \bibinfo{person}{Yeganeh
  Kordi}, \bibinfo{person}{Swaroop Mishra}, \bibinfo{person}{Alisa Liu},
  \bibinfo{person}{Noah~A. Smith}, \bibinfo{person}{Daniel Khashabi}, {and}
  \bibinfo{person}{Hannaneh Hajishirzi}.} \bibinfo{year}{2022}\natexlab{}.
\newblock \showarticletitle{Self-Instruct: Aligning Language Model with Self
  Generated Instructions}.
\newblock \bibinfo{journal}{\emph{ArXiv}}  \bibinfo{volume}{abs/2212.10560}
  (\bibinfo{year}{2022}).
\newblock


\bibitem[Wei et~al\mbox{.}(2021)]%
        {Wei2021FinetunedLM}
\bibfield{author}{\bibinfo{person}{Jason Wei}, \bibinfo{person}{Maarten Bosma},
  \bibinfo{person}{Vincent Zhao}, \bibinfo{person}{Kelvin Guu},
  \bibinfo{person}{Adams~Wei Yu}, \bibinfo{person}{Brian Lester},
  \bibinfo{person}{Nan Du}, \bibinfo{person}{Andrew~M. Dai}, {and}
  \bibinfo{person}{Quoc~V. Le}.} \bibinfo{year}{2021}\natexlab{}.
\newblock \showarticletitle{Finetuned Language Models Are Zero-Shot Learners}.
\newblock \bibinfo{journal}{\emph{ArXiv}}  \bibinfo{volume}{abs/2109.01652}
  (\bibinfo{year}{2021}).
\newblock


\bibitem[Xiao et~al\mbox{.}(2019)]%
        {Xiao2019HierarchicalNV}
\bibfield{author}{\bibinfo{person}{Teng Xiao}, \bibinfo{person}{Shangsong
  Liang}, {and} \bibinfo{person}{Zaiqiao Meng}.}
  \bibinfo{year}{2019}\natexlab{}.
\newblock \showarticletitle{Hierarchical Neural Variational Model for
  Personalized Sequential Recommendation}.
\newblock \bibinfo{journal}{\emph{The World Wide Web Conference}}
  (\bibinfo{year}{2019}).
\newblock


\end{thebibliography}

\appendix

\end{document}